\documentclass[a4paper]{jpconf}
\usepackage{graphicx}

\usepackage[frozencache=true]{minted}
\setminted[go]{frame=lines,fontsize=\footnotesize,tabsize=4}
\setminted[c++]{frame=lines,fontsize=\footnotesize,tabsize=4}

\begin{document}

\title{\texttt{Go-HEP}: writing concurrent software with ease and \texttt{Go}}

\author{S Binet$^1$}

\address{$^1$ Laboratoire de Physique de Clermont-Ferrand, CNRS/IN2P3, FR}

\ead{sebastien.binet@clermont.in2p3.fr}

\begin{abstract}
High Energy and Nuclear Physics (HENP) libraries are now required to be more and
more multi-thread-safe, if not multi-thread-friendly and multi-threaded.
This is usually done using the new constructs and library components offered by
the \texttt{C++11} and \texttt{C++14} standards.
These components are however quite low-level and hard to use and compose.

However, \texttt{Go} provides a set of better building
blocks for tackling concurrency: goroutines and channels.
This language is now used by the whole cloud industry: docker/moby, rkt,
Kubernetes are obvious flagships for \texttt{Go}.
But to be able to perform any meaningful physics analysis, one needs a set of
basic libraries (matrix operations, linear algebra, plotting, I/O, ...)
We present \texttt{Go-HEP}, a set of packages to easily write concurrent software to
interface with legacy HENP \texttt{C++} physics libraries.

\end{abstract}

\section{Introduction}

With the end of Dennard Scaling, High Energy and Nuclear Physics (HENP) libraries and applications are now required to be multi-thread safe.
This is to ensure the scaling of performances with new CPU architectures.
The new \texttt{C++} standards introduced new constructs and library components: \texttt{std::thread}, \texttt{std::mutex}, \texttt{std::lock}, \ldots
These components are however quite low-level and hard to use and compose, or easy to misuse.
Moreover, \texttt{C++} is still plagued with scalability issues.
Development speed is hindered by the interplay between the compilation model (based on \texttt{\#include}) and \texttt{C++} templates.
\texttt{C++} is a very large language, hard and subtle to teach and understand: this has an impact on code maintenance and the ability to bring newcomers up-to-speed on any given project.
On the user side, \texttt{C++} is showing deficiencies as well: installing a project's many dependencies can be quite difficult.
\texttt{Python} users will be surprised to learn there is still no \texttt{pip}-like mechanism to install, automatically and recursively, dependencies for a given project.
Finally, most HEP software stacks rely on applications dynamically linked to hundreds shared libraries: packaging and deployment can be time consuming tasks.

Fixing the whole \texttt{C++} ecosystem sounds like a rather daunting task.
It might be easier instead to start from a blank page, with a new language that addresses all the current deficiencies of \texttt{C++}, and improves the day-to-day life of a typical software programmer in a multi-core environment.

\texttt{Go}~\cite{ref-golang} was created in 2009 to address these issues.
We will first describe briefly the \texttt{Go} programming language, its concurrency primitives and the features of its tooling that make \texttt{Go} a great fit for HENP applications.
We will then introduce \texttt{Go-HEP}, a set of libraries and applications that aims to provide physicists with the basic tools to perform HENP analyses and integrate them in already existing \texttt{C++/Python} analysis pipelines.

\section{Introduction to \texttt{Go}}

\texttt{Go} is an open source language, released under the BSD-3 license in 2009.
It is a compiled language with a garbage collector and builtin support for reflection.
The language is reminiscent of \texttt{C/C++}: it shows a syntax similar to its older peers but adds first-class functions, closures and object-oriented programming via the concept of interfaces.
\texttt{Go} is already available, via the \texttt{gc} toolchain, on all major platforms (Linux, Windows, macOS, Android, iOS, \ldots) and for many architectures (\texttt{amd64}, \texttt{arm64}, \texttt{i386}, \texttt{s390x}, \texttt{mips64}, \ldots)
There is also a \texttt{Go} frontend to \texttt{GCC}, aptly named \texttt{gccgo}, that can can target all the platforms and architectures that the \texttt{GCC} toolchain supports.
Griesemer, Pike and Thompson created \texttt{Go} to replace the multi-threaded \texttt{C++} web servers that were hard to develop and slow to compile, even when using Google's massive infrastructure.
As such, \texttt{Go} exposes two builtin concurrency primitives: the \emph{goroutines} -- very lightweight green threads -- and the \emph{channels} -- typed conduits that connect goroutines together.


Launching goroutines is done by prepending a function (or method) call with the keyword \texttt{go}.
Goroutines can be described as green threads: very lightweight execution contexts that are multiplexed on native threads.
Each goroutine starts with a small stack (around $4KB$) that can grow and shrink as needed.
This allows to write applications with thousands of goroutines without the need to buy high-end servers.
Writing real world concurrent programs needs synchronization and communication: the ability to exchange data between goroutines.
This is achieved \emph{via} channels.
Not only do channels allow to exchange data in a type safe manner but they are also a synchronization point: a goroutine trying to send a token of data through a channel will block until there is another goroutine on the other end of the channel trying to extract data from the channel, and \emph{vice versa}.
The last piece that makes building concurrent programs in \texttt{Go} efficient is the keyword \texttt{select}.
This keyword is like a \texttt{switch} statement for controling the data flow between multiple channels and thus between multiple goroutines.
The concurrency builtin tools of \texttt{Go} are easier to reason about and more composable than mutexes and locks.

Moreover, \texttt{Go} comes with a package system and a builtin tool to build \texttt{Go} code.
There are no header files like in \texttt{C++} as they require to recursively process all dependent headers and thus slow the build.
Once compiled, \texttt{Go} packages are completely self-contained and do not require a client of package \texttt{p1} -- which itself depends on \texttt{p2} and \texttt{p3} -- to know anything about these packages.
The client only needs \texttt{p1}.
This greatly improves the scalability of the build process as well as its speed.
Thanks to the way third-party packages are imported in \texttt{Go}, \emph{e.g.} \mintinline{go}{import "github.com/pkg/errors"}, a build tool only needs to parse and understand \texttt{Go} code to discover (recursively) dependencies of a given package.
Coupled to the fact that package import strings contain the URL to the repository (GitHub, BitBucket, \ldots) holding the actual code, this allows a simple command -- \texttt{go get} -- to fetch code from the internet, (recursively) discover and fetch its dependencies and, finally build the whole artefact.
The instructions to install code are valid on all platforms that are supported by a given \texttt{Go} toolchain: one just needs to point \texttt{go get} at a repository.
Finally, \texttt{Go} code being very quick to compile makes static compilation manageable and enables simple deployment scenarii that boil down to \texttt{scp}-ing the resulting binary.
\texttt{Go} exposes a productive development environment that is concurrency friendly.
It is being used by many companies~\footnote{A non-exhaustive list is maintained here: \texttt{https://github.com/golang/go/wiki/GoUsers}.} beside Google, and in a variety of scenarii: from rocket telemetry to cloud systems to container orchestration.
But for \texttt{Go} to be useful in HENP, physicists need libraries and tools to carry analyses.
Moreover, these tools need to be interoperable with existing analyses pipelines.

\section{\texttt{Go-HEP}}

\texttt{Go-HEP} is a set of libraries and applications released under the BSD-3 license.
They allow High Energy and Nuclear Physicists to write efficient analysis code in the \texttt{Go} programming language.
The \texttt{go-hep.org/x/hep} organization provides a set of pure-\texttt{Go} packages and building blocks to:
\begin{itemize}
	\item write analyses in \texttt{Go},
	\item write data acquisition and monitoring systems in \texttt{Go},
	\item write control frameworks in \texttt{Go}.
\end{itemize}

As \texttt{Go-HEP} is written in pure-\texttt{Go}, the whole software suite and its dependencies are installable on all \texttt{Go} supported platforms (Linux, macOS, Windows, RPi3, \ldots) with:
\begin{verbatim}
 $> go get go-hep.org/x/hep/...
\end{verbatim}
The ellipsis (\ldots) at the end instructs the \texttt{go} tool to compile and install all the libraries and applications that are part of the \texttt{go-hep.org/x/hep} repository.

The libraries provided by \texttt{Go-HEP} can be broadly organized around two categories:
\begin{itemize}
	\item libraries that provide physics and statistical tools (Lorentz vectors, histograms and n-tuples, fits, jet clustering, fast detector simulation, plots, \ldots)
	\item libraries that provide low level interoperability with \texttt{C++} libraries, to allow \texttt{Go-HEP} users to integrate with existing analyses pipelines.
\end{itemize}

Indeed, analyses in HENP are mainly written in \texttt{C++} and \texttt{Python}.
Even though \texttt{Go} has a native foreign function interface called \texttt{cgo} that allows \texttt{Go} code to use \texttt{C} code and vice versa, the libraries under \texttt{go-hep.org/x/hep} do not call any \texttt{C++} library (via a \texttt{C} shim library.)
This decision allows to retain the quick edit-compile-run development cycle of \texttt{Go} and the easy deployment and cross-compilation of \texttt{Go} applications.
Instead, the strategy for interoperating with \texttt{C++} is to integrate with \emph{e.g.} ROOT~\cite{ref-ROOT} at the data file level.
\texttt{Go-HEP} provides read/write access for LCIO, LHEF, HepMC, SLHA and YODA files.
\texttt{Go-HEP} provides -- at the moment -- only read access to ROOT files via its \texttt{go-hep.org/x/hep/rootio} package.

In the following, we will describe two components of \texttt{Go-HEP}, \texttt{fads} and \texttt{hep/rootio}, that enable physics analyses.

\section{\texttt{fads}}

\subsection{\texttt{fads-app}}

\texttt{fads} is a "FAst Detector Simulation toolkit".
This library is built on top of \texttt{Go-HEP}'s control framework, \texttt{fwk}.
The control framework exposes a traditional API, with a \texttt{Task} type that can be started, process some event data and then stopped.
Each event is processed by a goroutine.
The framework runs these tasks in their own goroutine and lets the \texttt{Go} runtime deal with work stealing.
Data between tasks are exchanged via an event store service which itself utilizes \texttt{channel}s to ensure there are no data races.
Besides an event store service, \texttt{fwk} also provides services for message logging, histogramming and n-tupling. 
N-tuple and histograms (1D and 2D) are provided by the \texttt{hep/hbook} and \texttt{hep/hbook/ntup} packages and are persistified using \texttt{hep/rio}, a binary file format that heavily draws inspiration from \texttt{SIO}, the binary file format the LCIO community uses.

Data dependencies between tasks are described at the job configuration level: each task declares what are its inputs (the type of input and a name that identifies it) and its outputs.
\texttt{fwk} ensures there are no cycles between the tasks and connects the goroutines together via channels.
\texttt{fwk} enables task-level parallelism and event-level parallelism via the concurrency building blocks of \texttt{Go}.
For sub-task parallelism, users are required -- by construction -- to use the same building block, so everything is consistent and the \texttt{Go} runtime has the complete picture.

\texttt{fads} is itself a transliteration of Delphes~\cite{ref-delphes} (v3.0.12) but with \texttt{fwk} underneath instead of \texttt{ROOT}.
To assess the performances of \texttt{Go} and \texttt{fads} in a realistic settings, the whole ATLAS data card provided with Delphes has been implemented and packaged with \texttt{fads} under \texttt{hep/fads/cmd/fads-app}.
This application is composed of:
\begin{itemize}
\item a HepMC file reader task,
\item a particle propagator, a calorimeter simulator,
\item energy rescalers, momentum smearers,
\item electron, photon and muon isolation tasks,
\item b-tagging and tau-tagging tasks, and a jet-finder task.
\end{itemize}

\begin{figure}[h]
	\begin{center}
 \includegraphics[width=0.75\textwidth]{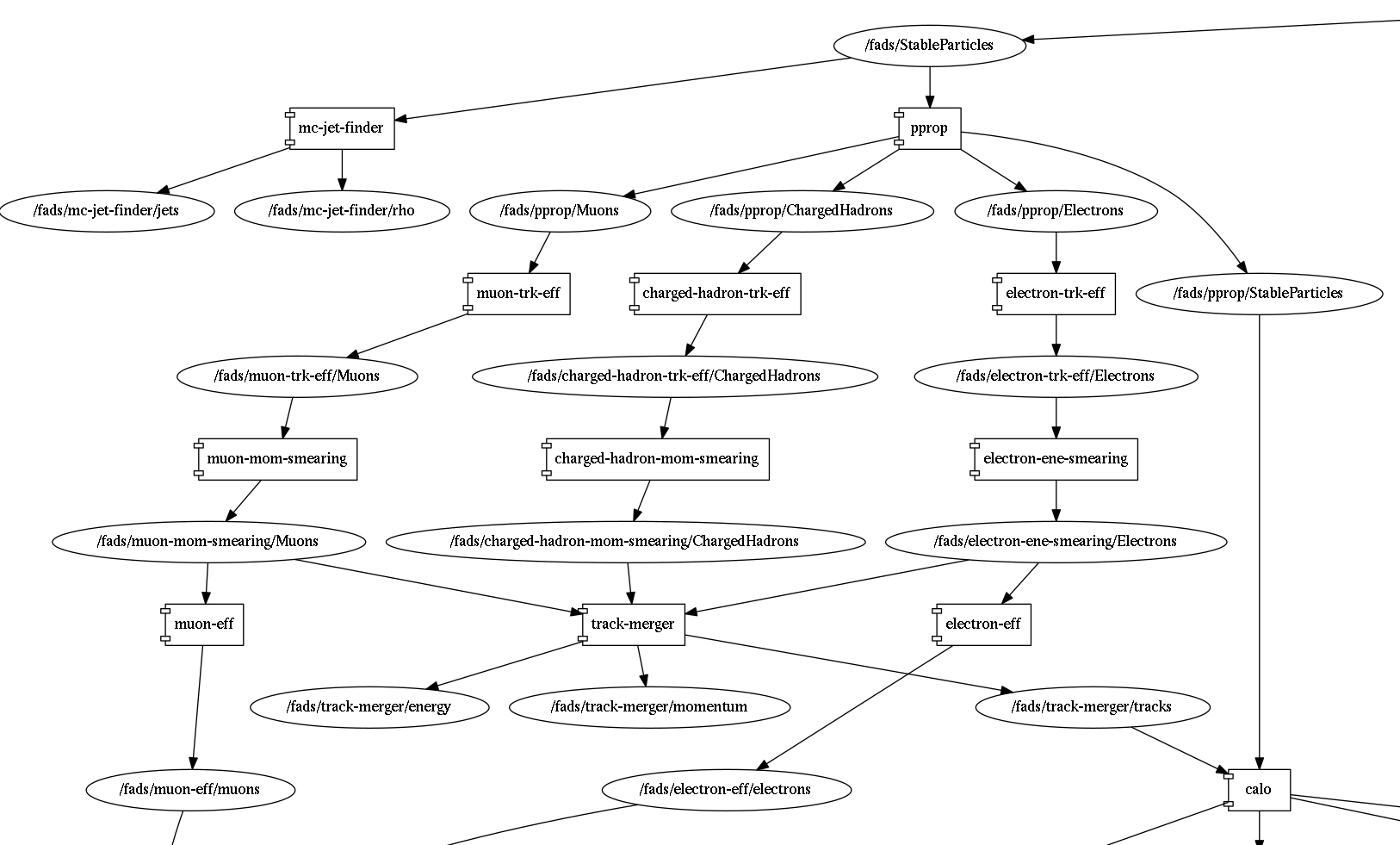}
	\end{center}
	\caption{\label{fig-dflow}Part of the directed acyclic graph of data dependencies between tasks composing the \texttt{fads-app}, a \texttt{Go} transliteration of the Delphes' ATLAS data card example. Rectangles are tasks, ellipses are data collections.}
\end{figure}

The jet finder task is based on a naive re-implementation of the \texttt{C++} FastJet~\cite{ref-fastjet} library: only the $N^3$ "dumb" clustering has been implemented so far.
A small part of the directed acyclic graph of the resulting \texttt{fads-app} application can be found in figure~\ref{fig-dflow}.

Delphes was compiled with \texttt{gcc-4.8} and the $N^3$ clustering strategy hard-coded, \texttt{fads} was compiled with \texttt{Go-1.9}.
Timings were obtained on a Linux Xeon CPU E5-4620@2.20GHz server with 64 cores and 128Gb RAM with a 10000 HepMC events input file.
\begin{figure}[h]
 \includegraphics[width=0.5\textwidth]{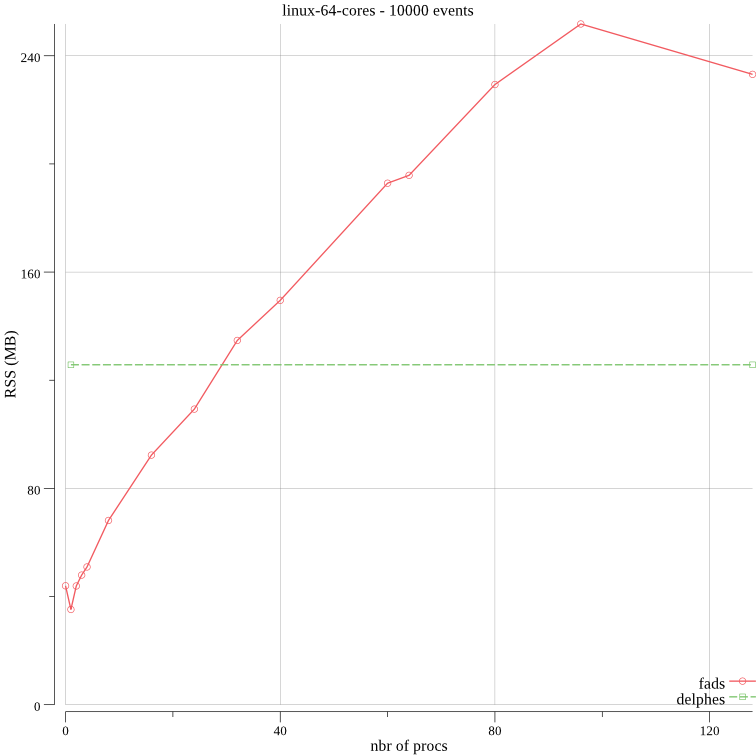}
 \includegraphics[width=0.5\textwidth]{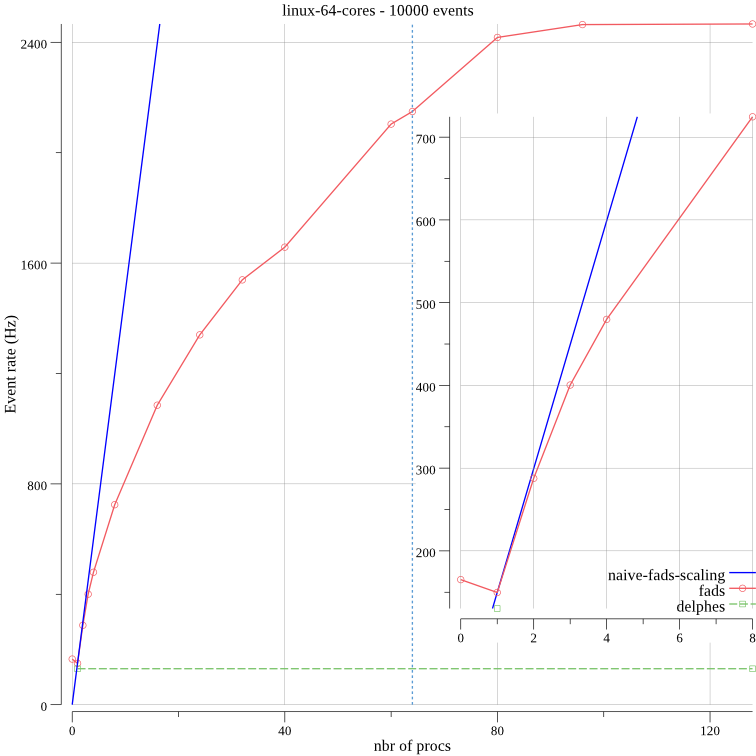}
	\caption{\label{fig-fads-perfs}Memory usage (left) and event processing rate (right) for \texttt{fads} (solid red curve) and \texttt{Delphes} (green dash line).}
\end{figure}

The results for various numbers of threads are shown in figure~\ref{fig-fads-perfs}.
As Delphes is not thread-safe, we only show the data for one Delphes process and then draw a flat line for visual aid.
The graph on the left shows a smaller memory footprint for \texttt{fads} that only matches that of Delphes' when 30 threads are used.
The graph on the right shows that \texttt{fads} achieves a better event processing rate even when using only one thread, or when using only one thread but with the sequential event loop instead of the concurrent event loop (data point for $n_{thread}=0$.)

\texttt{fads} achieves better performances than Delphes-3.0.12 on this data set.
The output simulated data for \texttt{fads} was matched bit-by-bit with that of Delphes up to the calorimetry stage~\footnote{We could not manage to match data further down the sequence because of Delphes' usage of PRNG to seed other PRNGs in the calorimeter. Control histograms agreed at the statistical level.}.
\texttt{fads} also achieves these performances without the need to merge output files.

\subsection{\texttt{fads-rivet-mc-generic}}

\texttt{fads} exports another application, \texttt{fads-rivet-mc-generic}, a reimplementation of the \texttt{MC\_GENERIC} analysis from the Rivet~\cite{ref-rivet} toolkit.

\begin{table}[h]
\begin{center}
  \begin{tabular}{ l | c | c | c | c }
	   & \texttt{MaxRSS} ($Mb$) & Real ($s$) & CPU ($s$) \\
    \hline
	   \texttt{Rivet} & 27 & 13.3 & 13.3 \\
	   \texttt{fads} & 23 & 5.7 & 5.7 \\
    \hline
  \end{tabular}
	\caption{\label{fig-rivet-perf-tab}Runtime performances of \texttt{Rivet} and \texttt{fads} to read a file with $Z$ events.}
\end{center}
\end{table}

\texttt{fads} shows better performances than the \texttt{C++} application.
As Rivet does not use ROOT for this application, the memory usage of Rivet is closer to that of \texttt{fads}.
However, the event processing rate of \texttt{fads} with only one thread is twice as good as the one of Rivet.
This is because the job steering and the event loop of Rivet are written in Python.

\section{\texttt{hep/rootio}}
The \texttt{hep/rootio} package is a pure-\texttt{Go} package that:
\begin{itemize}
\item decodes and understands the structure of \texttt{TFile}s, \texttt{TKey}s, \texttt{TDirectory} and \texttt{TStreamerInfo}s,
\item decodes and deserializes \texttt{TH1x}, \texttt{TH2x}, \texttt{TLeaf}, \texttt{TBranch} and \texttt{TTree}s. 
\end{itemize}
At the moment, \texttt{hep/rootio} only concerns itself with reading ROOT files, although writing ROOT files is on the roadmap.
Nonetheless, \texttt{hep/rootio} can already be useful and provides the following commands:
\begin{itemize}
	\item \texttt{cmd/root-ls} lists the content of ROOT files,
	\item \texttt{cmd/root-print} extracts histograms from ROOT files and saves them as PDF or PNG,
	\item \texttt{cmd/root-dump} prints the contents of \texttt{TTree}s, event by event,
	\item \texttt{cmd/root-diff} prints the differences between the contents of two \texttt{TTree}s,
	\item \texttt{root-cnv-npy} converts simple \texttt{TTree}s to \texttt{NumPy} array data files,
	\item \texttt{cmd/root-srv} allows to interactively plot histograms from a file and branches from a tree. \texttt{root-srv} being pure-\texttt{Go}, it can be hosted on Google AppEngine.
\end{itemize}

\texttt{rootio} can read flat \texttt{TTree}s with \texttt{C++} builtins, static and dynamic arrays of \texttt{C++} builtins.
\texttt{rootio} can also read \texttt{TTree}s with user defined classes containing \texttt{std::vector<T>}~\footnote{where \texttt{T} is a \texttt{C++} builtin or a \texttt{std::string} or a \texttt{TString}}, another user defined class, \texttt{std::string} or \texttt{TString}, arrays of \texttt{C++} builtins and \texttt{C++ builtins}.

To assess the performances of \texttt{hep/rootio} with regard to the original \texttt{C++} implementation, we created two input files of $10^6$ entries and 100 branches of \texttt{float64}.
One file was compressed with the default settings of ROOT-6.10 ($686 Mb$) and the other had no compression ($764 Mb$).

\begin{table}[h]
\begin{center}
  \begin{tabular}{ l | c | c | c | c }
	  \texttt{fCompression=0} & \texttt{VMem} ($Mb$) & \texttt{MaxRSS} ($Mb$) & Real ($s$) & CPU ($s$) \\
    \hline
	   \texttt{ROOT} & 517 & 258 & 6.7 & 6.6 \\
	   \texttt{hep/rootio} & 43 & 42 & 12.9 & 12.9 \\
    \hline
  \end{tabular}
\end{center}

\begin{center}
  \begin{tabular}{ l | c | c | c | c }
	  \texttt{fCompression=1} & \texttt{VMem} ($Mb$) & \texttt{MaxRSS} ($Mb$) & Real ($s$) & CPU ($s$) \\
    \hline
	   \texttt{ROOT} & 529 & 292 & 12.6 & 12.0 \\
	   \texttt{hep/rootio} & 83 & 82 & 35.8 & 35.8 \\
    \hline
  \end{tabular}
	\caption{\label{fig-rootio-perf-tab}Runtime performances of \texttt{C++/ROOT} and \texttt{hep/rootio} to read a file with $10^6$ events, using no compression (up) and with default compression (down).}
\end{center}
\end{table}

Table~\ref{fig-rootio-perf-tab} shows the results obtained running the two programs with the compressed file and the uncompressed file as inputs.
While the memory footprint of the \texttt{Go} program is almost an order of magnitude lower than its \texttt{C++} counterpart, it is twice as slow in the non-compressed case and almost thrice in the compressed case.
The degraded performances in the compressed case have been tracked back to the decompression package from the standard library which could probably be further optimized.
On the otherhand, the degraded performances in the non-compressed case come from the still young implementation of \texttt{hep/rootio} which could be also further optimized, \emph{e.g.} preemptively loading multiple consecutive entries.

\section{Conclusions}

\texttt{Go} improves on \texttt{C++/Python} and addresses their deficiences with regard to code distribution, code installation, compilation, development and runtime speeds.
\texttt{Go} also provides builtin facilities to tackle concurrency efficiently and easily.
\texttt{Go-HEP} provides some building blocks that are already competitive with battle-tested \texttt{C++} programs, both in terms of CPU, memory usage and cores' utilization.
Further improvements are still necessary in the ROOT I/O area, both in terms of performances and features, so that \texttt{Go-HEP} can be part of physics analyses pipelines.

\end{document}